\documentclass[a4paper]{article}
\usepackage{amscd,amsmath,amssymb}

\newcommand{\bT}{{\overline T}{}}

\newcommand{\bW}{{\overline W}{}}

\newcommand{\cN}{{ {\cal N}   }}

\newcommand{\tr}{\;\textrm{Tr}\;}

\newcommand{\be}{\begin{equation}}
\newcommand{\ee}{\end{equation}}
\newcommand{\bea}{\begin{eqnarray}}
\newcommand{\eea}{\end{eqnarray}}

\newcommand{\ba}{\begin{array}} \newcommand{\ea}{\end{array}}

\def\im{{\rm i}}

\newcommand{\nn}{\nonumber}
\def\theequation{\arabic{section}.\arabic{equation}}

\begin{document}
\thispagestyle{empty}
\vspace{2cm}
\begin{flushright}
\end{flushright}\vspace{2cm}
\begin{center}
{\Large\bf Superconformal mechanics from $\cN$-extended Euler-Calogero-Moser and Calogero models}
\end{center}
\vspace{1cm}
\begin{center}
{\Large\bf Sergey Krivonos${}^{a,b}$ and Anton Sutulin${}^a$
}
\end{center}

\begin{center}
{\it
${}^{a)}$Bogoliubov  Laboratory of Theoretical Physics, JINR,
141980 Dubna, Russia

${}^{b)}$Tomsk State University of Control Systems and Radioelectronics,
Lenin ave. 40, 634050 Tomsk, Russia}

\vspace{0.5cm}

{\tt   krivonos@theor.jinr.ru,sutulin@theor.jinr.ru }
\end{center}
\vspace{2cm}

\begin{abstract}
\noindent We investigate two-particle variants of the N-extended Calogero and Euler–Calogero–Moser models. Owing to translational invariance, the center-of-mass degrees of freedom and the associated fermions decouple, yielding specific systems of superconformal mechanics. Additional bosonic spin variables are straightforwardly incorporated into the supercharges and the Hamiltonian. For the Euler–Calogero–Moser model, the two-particle supersymmetry exhibits an unexpected extension from $Osp(N|2)$ to $SU(1,1|N)$ superconformal symmetry, whereas the Calogero model reduces to purely 
$Osp(N|2)$ superconformal mechanics. Finally, we demonstrate that introducing a larger number of fermions into the system provides a consistent pathway to overcoming well-known pathologies with higher supersymmetries inherent to these frameworks.
\end{abstract}


\vskip 0.5cm


\newpage
\setcounter{page}{1}
\setcounter{equation}{0}
\section{Introduction}
The enduring interest in superconformal mechanics is motivated by several key factors. At its core, superconformal mechanics is based on the supersymmetric extension of the one-dimensional conformal group, $SO(1,2)$. Consequently, the classification of the relevant $d=1, \cN$ supermultiplets and their interactions is crucial, having proven highly instrumental for understanding supersymmetry in various dimensions \cite{FT1, FT2, IKL, BIKL}. Furthermore, the AdS${}_2$/CFT${}_1$ correspondence establishes a holographic duality between AdS${}_2$ gravity backgrounds and their dual superconformal mechanics realizations \cite{S1, S2}. The high dimensionality of the internal space in AdS${}_2$ string solutions admits a vast landscape of configurations, as extensively illustrated in, e.g., \cite{Lozano1, Lozano2} and references therein. 

Several distinct frameworks have been developed to construct superconformal mechanics: the method of non-linear realizations \cite{nlr1, nlr2, nlr3}, the canonical Hamiltonian formalism \cite{KriNers, KriNers1, KriNers2}, and the superfield approach (see, e.g., \cite{SCM} and references therein). Although the literature on this subject continues to expand rapidly, the majority of existing studies focus on scenarios with $\cN \leq 8$ supersymmetries. In a different vein, the challenge of incorporating additional bosonic fields (beyond the dilaton) into superconformal mechanics has been addressed through various spin-variable extensions \cite{spin1, spin2, spin3}. While the origin of the pathologies associated with these extra bosonic variables in $SU(1,1|\cN)$ superconformal mechanics was thoroughly explored in \cite{Gal1}, satisfactory resolutions have only recently been proposed in \cite{KriNers2, Koz}.

Recalling the seminal work by de Alfaro, Fubini, and Furlan \cite{AFF}, their foundational setup can be derived from the two-particle Calogero model \cite{cal1} by decoupling the center-of-mass coordinate. Given that we now have access to $\cN$-extended supersymmetric variants of both the Euler–Calogero–Moser (ECM) \cite{KriS3} and the standard Calogero models \cite{KriS1, KriS2}, several natural questions arise: What is the precise structure of these models in the two-particle sector once the center-of-mass degrees of freedom are eliminated? Which supersymmetries are preserved in each case? Can additional bosonic spin-like variables be consistently integrated into these systems? Addressing these questions is the primary objective of the present paper.

The paper is structured as follows. In Section 2, we comprehensively examine the two-particle sector of the $\cN$-supersymmetric Euler–Calogero–Moser model. In Section 3, we uncover the hidden supersymmetry of this ECM model and construct its proper extensions, thereby resolving the issues related to additional bosonic variables. Section 4 is devoted to a detailed discussion of the superconformal mechanics emerging from the two-particle sector of the $\cN$-supersymmetric Calogero model, where the resulting dynamical symmetry is shown to be $Osp(\cN|2)$. Finally, we conclude with a brief discussion of our results.

\section{From Euler--Calogero--Moser model to superconformal mechanics}
\subsection{$\cN$-extended supersymmetric ECM model}

The bosonic Euler–Calogero–Moser (ECM) model~\cite{ECM1, ECM2} is defined by the Hamiltonian
\begin{equation}\label{bHam}
H = \frac{1}{2} \sum_{i=1}^n p_i^2 + \frac{1}{2} \sum_{i \neq j}^n \frac{\ell_{ij}^2}{(x_i - x_j)^2}.
\end{equation}
It depends on the coordinates $x_i(t)$ and canonical momenta $p_i(t)$ of each of the $n$ particles, as well as on the internal degrees of freedom encoded in the angular momenta $\ell_{ij} = -\ell_{ji}$. The coordinates and momenta satisfy the standard Poisson brackets
\begin{equation}\label{PB1}
\big\{ x_i, p_j \big\} = \delta_{ij},
\end{equation}
while the Poisson brackets of the angular momenta span the $\mathfrak{so}(n)$ algebra:
\begin{equation}\label{PB12}
\big\{ \ell_{ij}, \ell_{km} \big\} = \frac{1}{2} \big(\delta_{ik} \ell_{jm} + \delta_{jm} \ell_{ik} - \delta_{jk} \ell_{im} - \delta_{im} \ell_{jk}\big).
\end{equation}
The ECM model with the Hamiltonian~\eqref{bHam} possesses dynamical conformal invariance. Indeed, it is straightforward to verify that the dilatation generator $D$ and the conformal boost generator $K$, defined as
\begin{equation}\label{KD}
D = -\frac{1}{2} \sum_{i=1}^n x_i p_i \qquad \text{and} \qquad K = \frac{1}{2} \sum_{i=1}^n x_i^2,
\end{equation}
form, together with the Hamiltonian $H$~\eqref{bHam}, the one-dimensional conformal algebra $\mathfrak{so}(1,2)$:
\begin{equation}\label{cf1}
\big\{H, K\big\} = 2 D, \quad \big\{H, D \big\} = H, \quad \big\{ K,D\big\} = -K.
\end{equation}

The supersymmetric version of the ECM model with $\mathcal{N}$-extended super-Poincaré symmetry was constructed in~\cite{KriS3}. The supercharges $Q^a$ and $\bar{Q}_a$, which generate the $\mathcal{N}$-extended super-Poincaré algebra
\begin{equation}\label{NSP}
\big\{ Q^a , \bar{Q}_b \big\} = - 2 i \delta^a_b H, \quad \big\{ Q^a, Q^b \big\} = \big\{ \bar{Q}_a, \bar{Q}_b \big\} = 0, \quad a,b = 1, 2, \ldots, \mathcal{N}/2,
\end{equation}
take the form
\begin{equation}\label{QQb}
Q^a = \sum_{i=1}^n p_i \psi^a_i - \sum_{i \neq j}^n \frac{\left( \ell_{ij} + \Pi_{ij}\right) \rho^a_{ij}}{x_i - x_j} \quad \text{and} \quad
\bar{Q}_a = \sum_{i=1}^n p_i \bar{\psi}_{i\,a} - \sum_{i \neq j}^n \frac{\left( \ell_{ij} + \Pi_{ij}\right) \bar{\rho}_{ij\,a}}{x_i - x_j}.
\end{equation}
Here, the following fermionic variables are introduced:
\begin{itemize}
    \item $\mathcal{N} \times n$ fermions $\psi^a_i$ and $\bar{\psi}_{i\,a} = \left(\psi^a_i\right)^\dagger$ with $i = 1, \ldots, n$. These fermions can be combined with the bosonic coordinates $x_i(t)$ into $\mathcal{N}$ supermultiplets.
    \item $\mathcal{N} \times \frac{1}{2} n(n-1)$ additional fermions $\rho^a_{ij} = \rho^a_{ji}$ and $\bar{\rho}_{ij \,a} = \left(\rho^a_{ij}\right)^\dagger$, subject to the constraints $\rho^a_{ii} = \bar{\rho}_{ii\,a} = 0$ (no summation over repeated indices).
\end{itemize}
In total, the system utilizes $\mathcal{N} \times \frac{1}{2} n(n+1)$ fermions of types $\psi$ and $\rho$, which obey the brackets
\begin{align}\label{PB2}
\big\{ \psi^a_i, \bar{\psi}_{j\,b}\big\} &= -i \delta^a_b \delta_{ij}, \\
\big\{ \rho^a_{ij}, \bar{\rho}_{km\,b}\big\} &= -\frac{i}{2}\delta^a_b \big(1 - \delta_{ij}\big) \big(1 - \delta_{km}\big) \big( \delta_{ik}\delta_{jm} + \delta_{im}\delta_{jk}\big).
\end{align}
By employing these fermions, composite objects $\Pi_{ij}$ were introduced in~\cite{KriS3}:
\begin{equation}\label{Pi}
\Pi_{ij} = -\Pi_{ji} = - i \biggl[ \bigl( \psi^a_i - \psi^a_j\bigr) \bar{\rho}_{ij\,a} + \bigl( \bar{\psi}_{i\,a} - \bar{\psi}_{j\,a}\bigr) \rho^a_{ij} + \sum_{k=1}^n \left( \rho^a_{ik}\bar{\rho}_{kj\,a} - \rho^a_{jk}\bar{\rho}_{ki\,a}\right)\biggr],
\end{equation}
which satisfy the $\mathfrak{so}(n)$ Poisson brackets~\eqref{PB12}:
\begin{equation}\label{sonPi}
\big\{ \Pi_{ij}, \Pi_{km} \big\} = \frac{1}{2} \big( \delta_{ik} \Pi_{jm} + \delta_{jm} \Pi_{ik} - \delta_{jk} \Pi_{im} - \delta_{im} \Pi_{jk}\big).
\end{equation}
The supercharges $Q^a$ and $\bar{Q}_b$ form the $\mathcal{N}$-extended super-Poincaré algebra~\eqref{NSP} together with the full supersymmetric Hamiltonian:
\begin{equation}\label{Ham}
H = \frac{1}{2}\sum_{i=1}^n p_i^2 + \frac{1}{2} \sum_{i \neq j}^n \frac{\left( \ell_{ij} + \Pi_{ij}\right)^2 }{\left(x_i - x_j\right)^2}.
\end{equation}

\subsection{Two-Particle Sector and Center-of-Mass Decoupling}\label{subsec:2body_decoupling}

In the bosonic sector, translational invariance naturally leads to the decoupling of the center of mass of the ECM system. For the two-particle case ($i,j \in \{1,2\}$), one can introduce the center-of-mass and relative variables as follows:
\begin{equation}\label{nvars1}
X = x_1 + x_2, \quad x = x_1 - x_2, \qquad P = \frac{1}{2} \big(p_1 + p_2 \big), \quad p = \frac{1}{2} \big(p_1 - p_2 \big).
\end{equation}
In terms of these variables, the bosonic Hamiltonian~\eqref{bHam} decomposes into
\begin{equation}
H = \frac{1}{2} P^2 + \left( \frac{1}{2} p^2 + \frac{\ell_{12}^2}{x^2}\right) = H_{\text{free}} + H_{\text{cm}},
\end{equation}
where $H_{\text{free}} = \frac{1}{2} P^2$ and $H_{\text{cm}}$ dynamically commute, allowing one to disregard the free center-of-mass motion.

The reduction of the two-particle supersymmetric ECM system proceeds in a similar manner. First, we redefine the bosonic variables according to~\eqref{nvars1} and simultaneously introduce the collective and relative spinor variables:
\begin{equation}\label{nvars2}
\Psi^a = \psi^a_1 + \psi^a_2, \quad \bar{\Psi}_a = \bar{\psi}_{1 a} + \bar{\psi}_{2 a}, \qquad
\psi^a = \psi^a_1 - \psi^a_2, \quad \bar{\psi}_a = \bar{\psi}_{1 a} - \bar{\psi}_{2 a},
\end{equation}
along with the rescaled auxiliary fermions
\begin{equation}\label{nvars3}
\xi^a = 2 \rho^a_{12}, \qquad \bar{\xi}_a = 2 \bar{\rho}_{12 a}.
\end{equation}
Consequently, the full supercharges~\eqref{QQb} acquire the split form
\begin{equation}\label{SCharges1}
Q^a = P \Psi^a + \left( p \psi^a - \frac{1}{x} \left(g + \Pi\right) \xi^a\right), \qquad
\bar{Q}_a = P \bar{\Psi}_a + \left( p \bar{\psi}_a - \frac{1}{x} \left(g + \Pi\right) \bar{\xi}_a \right),
\end{equation}
where we have defined the coupling parameter and the composite fermionic current as
\begin{equation}\label{Pi1}
g = \ell_{12}, \qquad \Pi = \frac{i}{2} \big( \xi^a \bar{\psi}_a - \psi^a \bar{\xi}_a \big).
\end{equation}
The center-of-mass parts of the supercharges, $Q^a_{\text{free}} = P \Psi^a$ and $\bar{Q}_{\text{free } a} = P \bar{\Psi}_a$, satisfy the standard relation
\begin{equation}
\big\{ Q^a_{\text{free}}, \bar{Q}_{\text{free } b} \big\} = -4i \delta^a_b H_{\text{free}},
\end{equation}
providing a free supersymmetric extension of the center-of-mass degrees of freedom.

By focusing on the relative sector, we postulate the canonical non-vanishing Poisson brackets for the coordinates, relative, and auxiliary fermions:
\begin{equation}\label{PB}
\big\{ x, p \big\} = 1, \qquad \big\{ \xi^a, \bar{\xi}_b \big\} = - 2i \delta^a_b, \qquad \big\{ \psi^a, \bar{\psi}_b \big\} = - 2i \delta^a_b.
\end{equation}
Under these brackets, the isolated relative supercharges
\begin{equation}\label{SCharges2}
Q^a = p \psi^a - \frac{1}{x} \left(g + \Pi\right) \xi^a, \qquad \bar{Q}_a = p \bar{\psi}_a - \frac{1}{x} \left(g + \Pi\right) \bar{\xi}_a,
\end{equation}
span an $\mathcal{N}$-extended superconformal mechanics governed by the relative Hamiltonian $H$:
\begin{equation}\label{QQH}
\big\{ Q^a, Q^b \big\} = \big\{ \bar{Q}_a, \bar{Q}_b \big\} = 0, \qquad \big\{ Q^a, \bar{Q}_b \big\} = - 4i \delta^a_b H,
\end{equation}
where
\begin{equation}\label{rel_Ham}
H = \frac{1}{2} p^2 + \frac{1}{2x^2} \big( g + \Pi \big)^2.
\end{equation}

\subsection{Superconformal Properties}\label{subsec:superconformal_prop}

The structural properties of the supersymmetric Hamiltonian~\eqref{rel_Ham} imply the existence of an underlying dynamical superconformal symmetry. This invariance is generated by the conformal supercharges $S^a$ and $\bar{S}_a$, alongside the dilatation $D$ and conformal boost $K$, which are defined as follows:
\begin{equation}\label{OSp1}
S^a = x \psi^a, \quad \bar{S}_a = x \bar{\psi}_a, \qquad D = -\frac{1}{2} x p, \quad K = \frac{1}{2} x^2.
\end{equation}
The cross-brackets between the super-Poincaré charges $(Q, \bar{Q})$ and the superconformal charges $(S, \bar{S})$ play a critical role and close on the dilatation and $R$-symmetry generators:
\begin{equation}\label{QS1}
\begin{aligned}
\big\{ S^a, Q^b \big\} &= 2 \mathcal{T}^{ab}, & \big\{ S^a, \bar{Q}_b \big\} &= 2 \mathcal{V}^a{}_b + 4i \delta^a_b D, \\
\big\{ \bar{S}_a, Q^b \big\} &= -2 \mathcal{V}^b{}_a + 4i \delta^b_a D, \qquad & \big\{ \bar{S}_a, \bar{Q}_b \big\} &= 2 \bar{\mathcal{T}}_{ab},
\end{aligned}
\end{equation}
where the generators of the internal $R$-symmetry algebra take the form
\begin{equation}\label{OSp2}
\mathcal{V}^a{}_b = \frac{1}{2}\left( \xi^a \bar{\xi}_b + \psi^a \bar{\psi}_b\right), \quad
\mathcal{T}^{ab} = \frac{1}{2} (\xi^a \xi^b + \psi^a \psi^b), \quad
\bar{\mathcal{T}}_{ab} = \frac{1}{2} (\bar{\xi}_a \bar{\xi}_b + \bar{\psi}_a \bar{\psi}_b).
\end{equation}.

It is straightforward to verify that this $R$-symmetry sector closes into an $\mathfrak{so}(\mathcal{N})$ algebra:
\begin{align}\label{SO(N)}
\big\{ \mathcal{V}^a{}_b, \mathcal{V}^c{}_d \big\} &= i \delta^a_d \mathcal{V}^c{}_b - i \delta^c_b \mathcal{V}^a{}_d, \nonumber \\
\big\{ \mathcal{T}^{ab}, \bar{\mathcal{T}}_{cd} \big\} &= -i \delta^a_d \mathcal{V}^b{}_c + i \delta^a_c \mathcal{V}^b{}_d - i \delta^b_c \mathcal{V}^a{}_d + i \delta^b_d \mathcal{V}^a{}_c, \nonumber \\
\big\{ \mathcal{V}^a{}_b, \mathcal{T}^{cd} \big\} &= i \delta^d_b \mathcal{T}^{ac} - i \delta^c_b \mathcal{T}^{ad}, \quad
\big\{ \mathcal{V}^a{}_b, \bar{\mathcal{T}}_{cd} \big\} = i \delta^a_c \bar{\mathcal{T}}_{bd} - i \delta^a_d \bar{\mathcal{T}}_{bc}.
\end{align}

Specifically, the generators $\mathcal{V}^a{}_b$ span the $\mathfrak{u}(\mathcal{N}/2)$ subalgebra, while $\mathcal{T}^{ab}$ and $\bar{\mathcal{T}}_{cd}$ extend it to the full $\mathfrak{so}(\mathcal{N})$ structure. Consequently, the complete superconformal algebra realized by the system is identified as $\mathfrak{osp}(\mathcal{N}|2)$.

\setcounter{equation}0
\section{{$2\mathcal{N}$}-Extended Supersymmetry of the Two-Particle ECM Model}\label{sec:3}

\subsection{Extension of the Supersymmetry}\label{subsec:extension}

In the structure of the relative supercharges $Q^a$ and $\bar{Q}_a$ defined in~\eqref{SCharges2}, there exists an evident discrete symmetry that is disrupted by the simultaneous exchange of fermions $\psi^a \leftrightarrow \xi^a$ and $\bar{\psi}^a \leftrightarrow \bar{\xi}^a$. As a consequence of this invariance, one can introduce a set of novel supercharges:
\begin{equation}\label{SCharges3}
q^a = p \xi^a + \frac{1}{x} \left(g + \Pi\right) \psi^a, \qquad
\bar{q}_a = p \bar{\xi}_a + \frac{1}{x} \left(g + \Pi\right) \bar{\psi}_a,
\end{equation}
which form an isomorphic superalgebra closing on the same relative Hamiltonian~\eqref{rel_Ham}:
\begin{equation}\label{QQH3}
\big\{ q^a, q^b \big\} = \big\{ \bar{q}_a, \bar{q}_b \big\} = 0, \qquad \big\{ q^a, \bar{q}_b \big\} = - 4i \delta^a_b H.
\end{equation}
It is straightforward to verify that the hidden supercharges $\{q^a, \bar{q}_a\}$  commute with the original set $\{Q^a, \bar{Q}_a\}$:
\begin{equation}\label{QQqq}
\big\{ q^a, Q^b \big\} = \big\{ q^a, \bar{Q}_b \big\} = \big\{ \bar{q}_a, Q^b \big\} = \big\{ \bar{q}_a, \bar{Q}_b \big\} = 0.
\end{equation}
Consequently, the two-particle sector of the ECM system naturally admits a hidden $2\mathcal{N}$-extended super-Poincaré supersymmetry.

The internal $R$-symmetry group undergoes a corresponding extension. Upon introducing the new superconformal generators
\begin{equation}
s^a = x \xi^a, \qquad \bar{s}_a = x \bar{\xi}_a,
\end{equation}
the complete set of cross-brackets between the super-Poincaré and superconformal generators reads:
\begin{align}\label{new_conformal}
\big\{ Q^a, S^b \big\} &= -2 \mathcal{T}^{ab}, & \big\{ Q^a, \bar{S}_b \big\} &= -2 \mathcal{V}^a{}_b + 4i \delta^a_b D, \nonumber \\
\big\{ Q^a, s^b \big\} &= 2i \mathcal{W}^{ab}, & \big\{ Q^a, \bar{s}_b \big\} &= - 2i \mathcal{U}^a{}_b + 2i \delta^a_b (\tr \mathcal{U} + g), \nonumber \\
\big\{ \bar{Q}_a, S^b \big\} &= 2 \mathcal{V}^b{}_a + 4i \delta^b_a D, & \big\{ \bar{Q}_a, \bar{S}_b \big\} &= -2 \bar{\mathcal{T}}_{ab}, \nonumber \\
\big\{ \bar{Q}_a, s^b \big\} &= - 2i \mathcal{U}^b{}_a + 2i \delta^b_a (\tr \mathcal{U} + g), & \big\{ \bar{Q}_a, \bar{s}_b \big\} &= 2i \bar{\mathcal{W}}_{ab}, \nonumber \\
\big\{ q^a, S^b \big\} &= -2i \mathcal{W}^{ab}, & \big\{ q^a, \bar{S}_b \big\} &= 2i \mathcal{U}^a{}_b - 2i \delta^a_b (\tr \mathcal{U} + g), \nonumber \\
\big\{ q^a, s^b \big\} &= -2 \mathcal{T}^{ab}, & \big\{ q^a, \bar{s}_b \big\} &= -2 \mathcal{V}^a{}_b + 4i \delta^a_b D, \nonumber \\
\big\{ \bar{q}_a, S^b \big\} &= 2i \mathcal{U}^b{}_a - 2i \delta^b_a (\tr \mathcal{U} + g), & \big\{ \bar{q}_a, \bar{S}_b \big\} &= - 2i \bar{\mathcal{W}}_{ab}, \nonumber \\
\big\{ \bar{q}_a, s^b \big\} &= 2 \mathcal{V}^b{}_a + 4i \delta^b_a D, & \big\{ \bar{q}_a, \bar{s}_b \big\} &= -2 \bar{\mathcal{T}}_{ab}.
\end{align}
Here, the operators $\{\mathcal{V}^a{}_b, \mathcal{T}^{ab}, \bar{\mathcal{T}}_{ab}\}$ coincide with those defined in~\eqref{OSp2}, while the newly emerging currents $\{\mathcal{U}^a{}_b, \mathcal{W}^{ab}, \bar{\mathcal{W}}_{ab}\}$ are given by:
\begin{equation}\label{Gen}
\mathcal{U}^a{}_b = \frac{i}{2} \big(\xi^a \bar{\psi}_b - \psi^a \bar{\xi}_b \big), \quad
\mathcal{W}^{ab} = \frac{i}{2} \big(\xi^a \psi^b + \xi^b \psi^a\big), \quad
\bar{\mathcal{W}}_{ab} = \frac{i}{2} \big(\bar{\xi}_a \bar{\psi}_b + \bar{\xi}_b \bar{\psi}_a\big).
\end{equation}
The full set of generators comprises $\mathcal{N}^2/4$ currents $\mathcal{V}^a{}_b$, $\mathcal{N}^2/4$ currents $\mathcal{U}^a{}_b$, alongside $\mathcal{N}^2/4$ components for each of the $\mathcal{T}^{ab}$, $\mathcal{W}^{ab}$, $\bar{\mathcal{T}}_{ab}$, and $\bar{\mathcal{W}}_{ab}$ tensors. Together, they span the extended $\mathfrak{u}(\mathcal{N})$ $R-$-symmetry algebra.

\subsection{Supercharges Extended by Bosonic R-Symmetry Generators}\label{subsec:bosonic_R}

The integration of spin-like variables~\cite{spin1, spin2, spin3} into superconformal mechanics can be systematically achieved via a three-step procedure~\cite{KriNers}:
\begin{itemize}
    \item \textbf{Factorization:} Rewrite the part of supercharges as bilinear products of the purely fermionic realizations of the $R$-symmetry generators and the fundamental fermions.
    \item \textbf{Extension:} Deform the fermionic $R$-symmetry currents by adding their bosonic counterparts, constructed solely from commuting spin-like variables.
    \item \textbf{Closure Verification:} Check whether the deformed total currents satisfy the original superconformal algebra.
\end{itemize}

To implement this scheme, let us consider the original supercharges at vanishing coupling constant ($g=0$):
\begin{equation}\label{SCharges-non-g}
\begin{aligned}
Q^a &= p \psi^a - \frac{1}{x} \Pi \xi^a, \qquad & \bar{Q}_a &= p \bar{\psi}_a - \frac{1}{x} \Pi \bar{\xi}_a, \\
q^a &= p \xi^a + \frac{1}{x} \Pi \psi^a, \qquad & \bar{q}_a &= p \bar{\xi}_a + \frac{1}{x} \Pi \bar{\psi}_a.
\end{aligned}
\end{equation}
By employing the currents defined in~\eqref{OSp2} and~\eqref{Gen}, these charges can be identically re-factored into the following expressions:
\begin{align}\label{nQQH}
Q^a &= p \psi^a + \frac{1}{x} \left[ \tr \mathcal{U}_f \xi^a + (\mathcal{U}_f)^a{}_b \xi^b - i (\mathcal{V}_f)^a{}_b \psi^b - \mathcal{W}_f^{ab} \bar{\xi}_b - i \mathcal{T}_f^{ab} \bar{\psi}_b \right], \nonumber \\
\bar{Q}_a &= p \bar{\psi}_a + \frac{1}{x} \left[ \tr \mathcal{U}_f \bar{\xi}_a + (\mathcal{U}_f)^b{}_a \bar{\xi}_b + i (\mathcal{V}_f)^b{}_a \bar{\psi}_b - (\bar{\mathcal{W}}_f)_{ab} \xi^b + i (\bar{\mathcal{T}}_f)_{ab} \psi^b \right], \nonumber \\
q^a &= p \xi^a - \frac{1}{x} \left[ \tr \mathcal{U}_f \psi^a + (\mathcal{U}_f)^a{}_b \psi^b + i (\mathcal{V}_f)^a{}_b \xi^b - \mathcal{W}_f^{ab} \bar{\psi}_b + i \mathcal{T}_f^{ab} \bar{\xi}_b \right], \nonumber \\
\bar{q}_a &= p \bar{\xi}_a - \frac{1}{x} \left[ \tr \mathcal{U}_f \bar{\psi}_a + (\mathcal{U}_f)^b{}_a \bar{\psi}_b - i (\mathcal{V}_f)^b{}_a \bar{\xi}_b - (\bar{\mathcal{W}}_f)_{ab} \psi^b - i (\bar{\mathcal{T}}_f)_{ab} \xi^b \right].
\end{align}

We now introduce a set of bosonic, commuting spin-like variables $\{v^a, \bar{v}_b, w^a, \bar{w}_b\}$ satisfying the canonical Poisson brackets:
\begin{equation}\label{harm}
\big\{ v^a, \bar{v}_b \big\} = i \delta^a_b, \qquad \big\{ w^a, \bar{w}_b \big\} = -i \delta^a_b.
\end{equation}
The purely bosonic $R$-symmetry currents are then constructed as:
\begin{align}\label{bos-curr}
(\mathcal{V}_b)^a{}_b &= v^a \bar{v}_b - w^a \bar{w}_b, & (\mathcal{T}_b)^{ab} &= v^a w^b - v^b w^a, & (\bar{\mathcal{T}}_b)_{ab} &= -\bar{v}_a \bar{w}_b + \bar{v}_b \bar{w}_a, \nonumber \\
(\mathcal{U}_b)^a{}_b &= -v^a \bar{v}_b - w^a \bar{w}_b, & (\mathcal{W}_b)^{ab} &= v^a w^b + v^b w^a, & (\bar{\mathcal{W}}_b)_{ab} &= \bar{v}_a \bar{w}_b + \bar{v}_b \bar{w}_a.
\end{align}
The bosonic operators $\{\mathcal{V}_b, \mathcal{T}_b, \bar{\mathcal{T}}_b\}$ satisfy the same $\mathfrak{so}(\mathcal{N})$ structure equations as their fermionic counterparts~\eqref{SO(N)}.

Next, we define the total $R$-symmetry currents as the sum of the fermionic and bosonic realizations, e.g., $\mathcal{V}^a{}_b = (\mathcal{V}_f)^a{}_b + (\mathcal{V}_b)^a{}_b$ and $\mathcal{U}^a{}_b = (\mathcal{U}_f)^a{}_b + (\mathcal{U}_b)^a{}_b$. Substituting these combined currents into the factorized structures yields the fully extended supercharges:
\begin{align}\label{new-Q}
Q^a &= p \psi^a + \frac{1}{x} \left[ \tr (\mathcal{U}_f - \mathcal{U}_b) \xi^a + \mathcal{U}^a{}_b \xi^b - i \mathcal{V}^a{}_b \psi^b - \mathcal{W}^{ab} \bar{\xi}_b - i \mathcal{T}^{ab} \bar{\psi}_b \right], \nonumber \\
\bar{Q}_a &= p \bar{\psi}_a + \frac{1}{x} \left[ \tr (\mathcal{U}_f - \mathcal{U}_b) \bar{\xi}_a + \mathcal{U}^b{}_a \bar{\xi}_b + i \mathcal{V}^b{}_a \bar{\psi}_b - \bar{\mathcal{W}}_{ab} \xi^b + i \bar{\mathcal{T}}_{ab} \psi^b \right], \nonumber \\
q^a &= p \xi^a - \frac{1}{x} \left[ \tr (\mathcal{U}_f - \mathcal{U}_b) \psi^a + \mathcal{U}^a{}_b \psi^b + i \mathcal{V}^a{}_b \xi^b - \mathcal{W}^{ab} \bar{\psi}_b + i \mathcal{T}^{ab} \bar{\xi}_b \right], \nonumber \\
\bar{q}_a &= p \bar{\xi}_a - \frac{1}{x} \left[ \tr (\mathcal{U}_f - \mathcal{U}_b) \bar{\psi}_a + \mathcal{U}^b{}_a \bar{\psi}_b - i \mathcal{V}^b{}_a \bar{\xi}_b - \bar{\mathcal{W}}_{ab} \psi^b - i \bar{\mathcal{T}}_{ab} \xi^b \right].
\end{align}
The corresponding full quantum-mechanical Hamiltonian takes the form:
\begin{equation}\label{Ham-Currents}
H = \frac{1}{2} p^2 + \frac{1}{2x^2} \Big( 3 \tr(\mathcal{U}_b) \tr(\mathcal{U}_b) + 2 \tr(\mathcal{U}_b)\tr(\mathcal{U}_f) - \tr(\mathcal{U}_f) \tr(\mathcal{U}_f) - \mathcal{C} \Big),
\end{equation}
where the quadratic Casimir operator $\mathcal{C}$ of the combined algebra is defined by
\begin{equation}\label{Casimir}
\mathcal{C} = \tr(\mathcal{V}^2) + \tr(\mathcal{T} \bar{\mathcal{T}}) + \tr(\mathcal{U}^2) + \tr(\mathcal{W} \bar{\mathcal{W}}).
\end{equation}

To conclude this subsection, we note that the system admits a consistent reduction. One can explicitly verify the relations
\begin{equation}\label{red1}
\left\{ H, \psi^a + i \xi^a \right\}\Big\vert_{\text{red}} = \left\{ H, \bar{\psi}_a - i \bar{\xi}_a \right\}\Big\vert_{\text{red}} = 0,
\end{equation}
under the algebraic reduction constraints
\begin{equation}\label{red1a}
\xi^a \to i \psi^a, \quad \bar{\xi}_a \to -i \bar{\psi}_a, \quad w^a \to 0, \quad \bar{w}_a \to 0.
\end{equation}
Applying the reduction mapping~\eqref{red1a} to the modified supercharges~\eqref{new-Q} leads to:
\begin{align}\label{Qred}
Q_{\text{red}}^a &= p \psi^a - \frac{2i}{x} \left( (\mathcal{V}_f)^a{}_b + (\mathcal{V}_b)^a{}_b \right) \psi^b - \frac{i}{x} \tr \left( \mathcal{V}_f - \mathcal{V}_b \right) \psi^a, \nonumber \\
\bar{Q}_{\text{red } a} &= p \bar{\psi}_a + \frac{2i}{x} \left( (\mathcal{V}_f)^b{}_a + (\mathcal{V}_b)^b{}_a \right) \bar{\psi}_b + \frac{i}{x} \tr \left( \mathcal{V}_f - \mathcal{V}_b \right) \bar{\psi}_a,
\end{align}
where $(\mathcal{V}_f)^a{}_b = \psi^a \bar{\psi}_b$ and $(\mathcal{V}_b)^a{}_b = v^a \bar{v}b$. The reduced Hamiltonian matches exactly the model developed in~\cite{Koz}:
\begin{equation}\label{Hred}
H{\text{red}} = \frac{1}{2}p^2 - \frac{2}{x^2} (\mathcal{V}_b)^a{}_b (\mathcal{V}_f)^b{}_a + \frac{1}{2 x^2} \left(\tr(\mathcal{V}_f - \mathcal{V}_b)\right)^2.
\end{equation}

\setcounter{equation}0
\section{Superconformal Mechanics from the Supersymmetric Calogero Model}\label{sec:4}

\subsection{$\mathcal{N}$-Extended Supersymmetric Calogero Model}\label{subsec:Calogero_model}

To construct the $\mathcal{N}$-supersymmetric extension of the two-particle Calogero model, we introduce the standard phase space equipped with fermionic variables~\cite{KriS1, KriS2}:
\begin{itemize}
    \item The particle coordinates $x_1, x_2$ and their canonically conjugate momenta $p_1, p_2$, satisfying the Poisson brackets:
    \begin{equation}
    \big\{ x_i, p_j \big\} = \delta_{ij}.
    \end{equation}
    \item $4\mathcal{N}$ fermionic variables $\rho^a_{ij}$ and $\bar{\rho}_{a\, ij}$ (where $a = 1, 2, \ldots, \mathcal{N}/2$ and $i, j \in \{1,2\}$), obeying the canonical brackets:
    \begin{equation}
    \big\{ \rho^a_{ij}, \bar{\rho}_{b\, kl} \big\} = -i \delta^a_b \delta_{il} \delta_{jk}.
    \end{equation}
\end{itemize}

Using these fundamental variables, we define the composite operators
\begin{equation}\label{Pi2}
\Pi_{ij} = \sum_{a=1}^{\mathcal{N}/2} \sum_{k=1}^2 \left( \rho^a_{ik} \bar{\rho}_{a\, kj} + \bar{\rho}_{a\, ik} \rho^a_{kj} \right),
\end{equation}
which span the standard $\mathfrak{su}(2)$ algebra:
\begin{equation}\label{su(n)}
\big\{ \Pi_{ij} , \Pi_{km} \big\} = i \left( \delta_{im} \Pi_{kj} - \delta_{kj}\Pi_{im} \right).
\end{equation}

The supercharges $Q^a$ and $\bar{Q}_a$, which generate the $\mathcal{N}$-extended super-Poincaré algebra
\begin{equation}
\big\{Q^a, \bar{Q}_b\big\} = -2i \delta^a_b H, \qquad \big\{ Q^a, Q^b \big\} = \big\{ \bar{Q}_a, \bar{Q}_b \big\} = 0,
\end{equation}
take a remarkably compact form:
\begin{equation}\label{QQB}
Q^a = \sum_{i=1}^2 p_i \rho^a_{ii} - i \sum_{i \neq j}^2 \frac{\left(g + \Pi_{jj} - \Pi_{ij}\right) \rho^a_{ji}}{x_i - x_j}, \qquad 
\bar{Q}_a = \sum_{i=1}^2 p_i \bar{\rho}_{a\, ii} + i \sum_{i \neq j}^2 \frac{\left(g + \Pi_{ii} - \Pi_{ji}\right) \bar{\rho}_{a\, ij}}{x_i - x_j},
\end{equation}
while the corresponding total supersymmetric Hamiltonian reads
\begin{equation}\label{H}
H = \frac{1}{2} \sum_{i=1}^2 p_i^2 + \frac{1}{2} \sum_{i \neq j}^2 \frac{\left(g + \Pi_{jj} - \Pi_{ij}\right)\left(g + \Pi_{ii} - \Pi_{ji}\right)}{\left( x_i - x_j\right)^2}.
\end{equation}

A distinctive feature of this formulation lies in the non-trivial involution properties of the fermions under Dirac conjugation~\cite{KriS2}:
\begin{equation}\label{conjugation}
\left(\rho^a_{ij}\right)^\dagger = \frac{g + \Pi_{jj}}{g + \Pi_{ii}} \bar{\rho}_{a\, ji}, \qquad \text{and} \qquad \left(\Pi_{ij}\right)^\dagger = \frac{g + \Pi_{jj}}{g + \Pi_{ii}} \Pi_{ji}.
\end{equation}
With respect to these conjugation rules, the supercharges $Q^a$ and $\bar{Q}_a$ are properly conjugate to each other, and the Hamiltonian $H$~\eqref{H} remains strictly self-adjoint within the $\mathcal{N}$-extended Calogero framework.

\subsection{Center-of-Mass Decoupling and Superconformal Mechanics}\label{subsec:Calogero_decoupling}

In analogy with the ECM model, translational invariance ensures that the center-of-mass degrees of freedom completely decouple from the relative motion. We perform this separation by introducing the collective variables $X$ and $P$, along with the center-of-mass fermions $\Psi^a$ and $\bar{\Psi}_a$:
\begin{equation}\label{nvars22}
X = x_1 + x_2, \quad P = \frac{1}{2}\big(p_1 + p_2 \big), \qquad \Psi^a = \frac{1}{2}\left(\rho^a_{11} + \rho^a_{22}\right), \quad \bar{\Psi}_a = \frac{1}{2}\left( \bar{\rho}_{a\,11} + \bar{\rho}_{a\,22}\right).
\end{equation}
The isolated center-of-mass supercharges $Q^a_{\text{free}}$ and $\bar{Q}_{\text{free } a}$ close on the free Hamiltonian $H_{\text{free}} = \frac{1}{2} P^2$, satisfying the standard super-Poincaré algebra:
\begin{equation}\label{nvars11}
\big\{ Q^a_{\text{free}}, \bar{Q}_{\text{free } b} \big\} = -i \delta^a_b H_{\text{free}}, \qquad \big\{Q^a_{\text{free}}, Q^b_{\text{free}}\big\} = \big\{\bar{Q}_{\text{free } a}, \bar{Q}_{\text{free } b}\big\} = 0.
\end{equation}

After projecting out the center-of-mass dynamics, the relative supercharges emerging from~\eqref{QQB} take the form (where $x = x_1 - x_2$ and $p = \frac{1}{2}(p_1 - p_2)$):
\begin{align}\label{QQB1}
Q^a &= \frac{1}{2} p \left(\rho^a_{11} - \rho^a_{22}\right) + \frac{i}{2 x} \left[ -\left(\Pi_{22} - \Pi_{12}\right) \rho^a_{21} + \left(\Pi_{11} - \Pi_{21}\right)\rho^a_{12} + g \left(\rho^a_{12} - \rho^a_{21}\right) \right], \nonumber \\
\bar{Q}_a &= \frac{1}{2} p \left(\bar{\rho}_{a\,11} - \bar{\rho}_{a\,22}\right) - \frac{i}{2 x} \left[ -\left(\Pi_{11} - \Pi_{21}\right) \bar{\rho}_{a\,12} + \left(\Pi_{22} - \Pi_{12}\right)\bar{\rho}_{a\, 21} - g\left(\bar{\rho}_{a\,12} - \bar{\rho}_{a\, 21}\right) \right],
\end{align}
and the corresponding relative Hamiltonian becomes
\begin{equation}\label{H1}
H = \frac{1}{2} p^2 + \frac{1}{2 x^2} \left(g + \Pi_{22} - \Pi_{12}\right) \left(g + \Pi_{11} - \Pi_{21}\right).
\end{equation}

The relative system governed by~\eqref{QQB1} and~\eqref{H1} possesses a dynamic $\mathfrak{osp}(\mathcal{N}|2)$ superconformal symmetry. To make this manifest, we introduce the generators of the $\mathfrak{so}(\mathcal{N})$ $R$-symmetry algebra:
\begin{align}\label{fRsym}
(\mathcal{T}_f)^{ab} &= \frac{1}{2}\left( \rho^a_{11} - \rho^a_{22}\right) \left( \rho^b_{11} - \rho^b_{22}\right) + \rho^a_{12}\rho^b_{21} - \rho^b_{12}\rho^a_{21}, \nonumber \\
(\mathcal{V}_f)^a{}_b &= \frac{1}{2}\left( \rho^a_{11} - \rho^a_{22}\right) \left( \bar{\rho}_{b\,11} - \bar{\rho}_{b\, 22}\right) + \rho^a_{12}\bar{\rho}_{b\, 21} + \rho^a_{21}\bar{\rho}_{b\, 12}, \nonumber \\
(\bar{\mathcal{T}}_f)_{ab} &= \frac{1}{2}\left( \bar{\rho}_{a\,11} - \bar{\rho}_{a\,22}\right) \left( \bar{\rho}_{b\,11} - \bar{\rho}_{b\, 22}\right) + \bar{\rho}_{a\,12}\bar{\rho}_{b\,21} - \bar{\rho}_{b\,12}\bar{\rho}_{a\,21}.
\end{align}
Defining the superconformal generator partners as
\begin{equation}
\mathcal{S}^a = x \left(\rho^a_{11} - \rho^a_{22}\right), \qquad \bar{\mathcal{S}}_a = x \left(\bar{\rho}_{a\,11} - \bar{\rho}_{a\,22}\right),
\end{equation}
one finds that their cross-brackets with the super-Poincaré charges close directly on the dilatation $D = -\frac{1}{2}xp$ and the $R$-symmetry currents:
\begin{equation}
\begin{aligned}
\big\{\mathcal{S}^a, Q^b\big\} &= (\mathcal{T}_f)^{ab}, \qquad & \big\{\mathcal{S}^a, \bar{Q}_b\big\} &= 4i \delta^a_b D + (\mathcal{V}_f)^a{}_b, \\
\big\{Q^a, \bar{\mathcal{S}}_b\big\} &= 4i \delta^a_b D - (\mathcal{V}_f)^a{}_b, \qquad & \big\{\bar{\mathcal{S}}_a, \bar{Q}_b\big\} &= (\bar{\mathcal{T}}_f)_{ab}.
\end{aligned}
\end{equation}
As required for the full algebraic consistency, all $R$-symmetry currents are dynamically conserved quantities:
\begin{equation}
\big\{H, (\mathcal{V}_f)^a{}_b\big\} = 0, \qquad \big\{H, (\mathcal{T}_f)^{ab}\big\} = 0, \qquad \big\{H, (\bar{\mathcal{T}}_f)_{ab}\big\} = 0.
\end{equation}

Furthermore, the relative supercharges~\eqref{QQB1} can be explicitly re-factored in terms of the fermionic $R$-symmetry generators~\eqref{fRsym}:
\begin{align}\label{QQB2}
Q^a &= \frac{1}{2} p\left( \rho^a_{11} - \rho^a_{22}\right) - \frac{i}{2 x}\left[ (\mathcal{V}_f)^a{}_b(\rho^b_{11} - \rho^b_{22}) + (\mathcal{T}_f)^{ab}(\bar{\rho}_{b\,11} - \bar{\rho}_{b\, 22})\right] + \frac{i}{x} \mathcal{Z}_f(\rho^a_{12} + \rho^a_{21}) \nonumber \\
& \quad + \frac{ig}{2x}(\rho^a_{12} - \rho^a_{21}), \nonumber \\
\bar{Q}_a &= \frac{1}{2} p\left( \bar{\rho}_{a\,11} - \bar{\rho}_{a\,22}\right) + \frac{i}{2 x}\left[ (\mathcal{V}_f)^b{}_a(\bar{\rho}_{b\,11} - \bar{\rho}_{b\,22}) - (\bar{\mathcal{T}}_f)_{ab}(\rho^b_{11} - \rho^b_{22})\right] + \frac{i}{x} \mathcal{Z}_f(\bar{\rho}_{a\,12} + \bar{\rho}_{a\,21}) \nonumber \\
& \quad + \frac{ig}{2x}(\bar{\rho}_{a\,12} - \bar{\rho}_{a\,21}),
\end{align}
where the auxiliary internal operator $\mathcal{Z}_f$ is defined as
\begin{equation}\label{fZ}
\mathcal{Z}_f = \rho^a_{12}\bar{\rho}_{a\,21} - \rho^a_{21}\bar{\rho}_{a\,12},
\end{equation}
and commutes with all $\mathfrak{so}(\mathcal{N})$ generators~\eqref{fRsym}.

The corresponding bosonic $R$-symmetry sector is realized by introducing the commuting variables $\{v^a, \bar{v}_b, w^a, \bar{w}_b\}$ from~\eqref{harm}, forming the currents:
\begin{equation}\label{bosR}
(\mathcal{V}_b)^a{}_b = v^a \bar{v}_b - w^a \bar{w}_b, \quad (\mathcal{T}_b)^{ab} = v^a w^b - v^b w^a, \quad (\bar{\mathcal{T}}_b)_{ab} = \bar{v}_a \bar{w}_b - \bar{v}_b \bar{w}_a, \quad \mathcal{Z}_b = \frac{1}{2}\left(v^a \bar{v}_a + w^a \bar{w}_a\right).
\end{equation}
We can now extend the full supercharges by coupling these bosonic spin variables to the fundamental fermions. Demanding closure under the super-Poincaré algebra fixes the modified total supercharges to the following form:
\begin{align}\label{QQB3}
\mathbb{Q}^a &= Q^a + \frac{i}{2 x}\left( (\mathcal{V}_b)^a{}_b \eta^b - (\mathcal{T}_b)^{ab}\bar{\eta}_b \right) - i \frac{\hat{g}}{x} \mathcal{Z}_b\left(\rho^a_{12} - \rho^a_{21}\right), \nonumber \\
\bar{\mathbb{Q}}_a &= \bar{Q}_a + \frac{i}{2 x} \left( (\mathcal{V}_b)^b{}_a \bar{\eta}_b - (\bar{\mathcal{T}}_b)_{ab}\eta^b \right) - i \frac{\hat{g}}{x} \mathcal{Z}_b\left(\bar{\rho}_{a\,12} - \bar{\rho}_{a\,21}\right),
\end{align}
where the collective fermionic combinations $\eta^a$ and $\bar{\eta}_a$ are given by
\begin{equation}
\eta^a = \rho^a_{11} - \rho^a_{22} + \rho^a_{12} - \rho^a_{21}, \qquad \bar{\eta}_a = \bar{\rho}_{a\,11} - \bar{\rho}_{a \,22} + \bar{\rho}_{a\,12} - \bar{\rho}_{a\,21}.
\end{equation}
The resulting comprehensive extended Hamiltonian evaluates to
\begin{align}
H &= \frac{1}{2} p^2 + \frac{1}{2 x^2} \left(g + 2 \hat{g} \mathcal{Z}_b + \Pi_{22} - \Pi_{12}\right)\left(g - 2 \hat{g} \mathcal{Z}_b + \Pi_{11} - \Pi_{21}\right) \nonumber \\
& \quad + \frac{1}{2 x^2}\left( \bar{w}_a v_b \eta^a \eta^b + v_a w_b \bar{\eta}^a \bar{\eta}^b + \bar{w}_a w_b \eta^a \bar{\eta}^b - \bar{v}_a v_b \eta^a \bar{\eta}^b \right).
\end{align}
Finally, in full analogy with the ECM system, one can directly establish the consistency of a dynamic algebraic reduction:
\begin{equation}
\left\{ H, \rho^a_{21} + \rho^a_{12}\right\} \Big\vert_{\text{red}} = 0, \qquad \left\{ H, \bar{\rho}_{a,21} + \bar{\rho}_{a,12}\right\}\Big\vert_{\text{red}} = 0,
\end{equation}
where $\vert_{\text{red}}$ denotes the constraint hyperplane:
\begin{equation}
\Big\vert_{\text{red}} = \Big\{ \rho^a_{21} \to -\rho^a_{12}, \quad 
\bar{\rho}_{a,12} \to -\bar{\rho}_{a,21} \Big\}.
\end{equation}
This reduction mapping provides a continuous trajectory connecting the extended Calogero supercharges~\eqref{QQB3} to the specific structures of the ECM model, which can be further truncated to the standard \(\mathcal{N}\)-extended systems.

\setcounter{equation}0
\section{Conclusion and Discussion}\label{sec:concl}

To summarize, in this paper we have investigated the two-particle sectors of the $\mathcal{N}$-extended Euler–Calogero–Moser (ECM) and Calogero models. Owing to translational invariance, the center-of-mass degrees of freedom and the corresponding collective fermions completely decouple, yielding specific systems of superconformal mechanics. Our analysis was primarily based on the extensive capabilities offered by the canonical Hamiltonian formalism. This choice of framework was motivated chiefly by the current lack of an alternative off-shell superfield description for systems featuring higher supersymmetries with $\mathcal{N} > 8$. Another notable advantage of the Hamiltonian approach is its facility for incorporating additional bosonic variables (spin variables) directly into the supercharges and the Hamiltonian. This is achieved by utilizing bosonic currents that realize the $R$-symmetry of the underlying superconformal algebra, a technique whose utility has previously been demonstrated for $\mathcal{N}=8$ configurations in~\cite{KriNers, KriNers1}. These models provide a precise microscopic description for the world-line dynamics of superparticles moving in a near-horizon extremal black hole geometry (AdS${}_2 \times S^d$). They serve as the CFT${}_{1}$ dual candidates in the (AdS${}_2/$CFT${}_1$) correspondence, where higher supersymmetries match the extensive isometry groups of string theory compactifications.

The superconformal mechanics systems constructed herein exhibit several intriguing peculiarities. For the two-particle Euler–Calogero–Moser model, the supersymmetry admits an unexpected enhancement from $\mathfrak{osp}(\mathcal{N}|2)$ to the larger $\mathfrak{su}(1,1|\mathcal{N}/2)$ superconformal symmetry. Conversely, the reduction of the Calogero model results strictly in purely $\mathfrak{osp}(\mathcal{N}|2)$ superconformal mechanics. 

It is well known that the construction of superconformal mechanics featuring $\mathfrak{osp}(\mathcal{N}|2)$ symmetry with $\mathcal{N} > 8$ typically encounters significant structural pathologies~\cite{Cher1}. A consistent pathway to overcoming these pathologies associated with spin-like variables is to intentionally increase the number of fermions in the system. Thus, to realize a well-behaved superconformal mechanics with $\mathfrak{osp}(\mathcal{N}|2)$ symmetry for $\mathcal{N} > 8$, one must employ an expanded set of fermionic degrees of freedom compared to the minimal requirements predicted by the standard $\mathcal{N}$-extended frameworks. Interestingly, the situation for the $\mathfrak{su}(1,1|\mathcal{N}/2)$ superconformal mechanics proves more straightforward: it is sufficient to realize the bosonic $R$-symmetry currents in terms of semi-dynamical variables~\cite{spin1, spin2, spin3, KriNers} and subsequently utilize these currents to deform the initial supercharges.  The successfully generated interaction terms involving spin variables provide a concrete benchmark for developing new ($d=1$) superfield techniques and off-shell formulations for $\mathcal{N} > 8$, which remain largely unmapped.

\section*{Declaration of competing interest}
The authors declare that they have no known competing financial interests or personal relationships that could have appeared to influence the
work reported in this paper.

\section*{Acknowledgements}
This work of S.K. was partially supported by the Russian Science Foundation, grant No 23-11-00002-Ext.

\section*{Data availability}
No data was used for the research described in the article.

\def\theequation{A.\arabic{equation}}
\setcounter{equation}0
\section*{Appendix A}
The structure relations of the $U(\cN)$ algebra are
\bea\label{UN}
&& 
\big\{ V^a{}_b, V^c{}_d \big\} = \im \delta^a_d V^c{}_b -\im \delta^c_b V^a{}_d, \qquad a,b,c,d = 1,\ldots, \frac{\cN}{2} \nn \\
&&\big\{ V^a{}_b, T^{cd} \big\} = \im \delta^d_b T^{ac} -\im \delta^c_b T^{ad}, \quad
\big\{ V^a{}_b, \bT_{cd} \big\} = \im \delta^a_c \bT_{bd} -\im \delta^a_d \bT_{bc}, \nn\\
&& 
\big\{ T^{ab}, \bT_{cd} \big\} =- \im \delta^a_d V^b{}_c +\im \delta^a_c V^b{}_d - \im \delta^b_c V^a{}_d +-\im \delta^b_d V^a{}_c, \nn\\
&& 
\big\{ V^a{}_b, W^{cd} \big\} = - \im \delta^c_b W^{ad} - \im \delta^d_b W^{ac}, \quad \big\{ T^{ab}, W^{cd} \big\} = 0, \nn\\
&& 
\big\{ W^{ab},  \bT_{cd} \big\} = -\im \delta^a_d U^b{}_c + \im \delta^a_c U^b{}_d - \im \delta^b_d U^a{}_c + \im \delta^b_c U^a{}_d, \nn\\
&& 
\big\{ V^a{}_b, \bW_{cd} \big\} = \im \delta^a_c \bW_{bd} + \im \delta^a_d \bW_{bc}, \quad \big\{ \bT_{ab}, \bW_{cd} \big\} = 0, \nn\\
&& 
\big\{ T^{ab}, \bW_{cd} \big\} = - \im \delta^a_d U^b{}_c + \im \delta^b_d U^a{}_c - \im \delta^a_c U^b{}_d + \im \delta^b_c U^a{}_d, \nn\\
&&
\big\{ V^a{}_b, U^c{}_d \big\} = \im \delta^a_d U^c{}_b -\im \delta^c_b U^a{}_d, \nn\\
&& \big\{ U^a{}_b, T^{cd} \big\} = \im \delta^c_b W^{ad} - \im \delta^d_b W^{ac}, \quad
\big\{ U^a{}_b, \bT_{cd} \big\} = -\im \delta^a_d \bW_{bc} + \im \delta^a_c \bW_{bd}, \nn\\
&& 
\big\{ U^a{}_b, U^c{}_d \big\} = \im \delta^a_d V^c{}_b -\im \delta^c_b V^a{}_d, \nn\\
&&\big\{ U^a{}_b, W^{cd} \big\} = \im \delta^d_b T^{ac} + \im \delta^c_b T^{ad}, \quad
\big\{ U^a{}_b, \bW_{cd} \big\} = \im \delta^a_c \bT_{bd} + \im \delta^a_d \bT_{bc}, \nn\\
&& 
\big\{ W^{ab}, \bW_{cd} \big\} = - \im \delta^a_d V^b{}_c -\im \delta^a_c V^b{}_d - \im \delta^b_c V^a{}_d -\im \delta^b_d V^a{}_c.
\eea

\end{document}